\documentclass[superscriptaddress,showkeys,letterpaper]{revtex4-1}
\usepackage{amssymb,amsmath,amsfonts}
\title{Bound states and scattering coefficients of self-adjoint Hamiltonians with a mass jump}
\begin{document}
\title{Bound states and scattering coefficients of self-adjoint Hamiltonians with a mass jump}
\author{L. A. Gonz\'{a}lez - D\'{\i}az.}
\email{lgonzale@ivic.gob.ve}
\affiliation{Laboratorio de Din\'amica Estoc\'astica, Centro de F\'{\i}sica, Instituto Venezolano de Investigaciones Cient\'{\i}ficas, Caracas 1020 - A,
Venezuela.}
\affiliation{Centro de Investigaciones de Matem\'atica y F\'{\i}sica, Departamento de Matem\'atica y F\'{\i}sica, Instituto Pedag\'ogico de Caracas, UPEL, Av. P\'aez, Caracas 1021, Venezuela.}
\author{S. D\'{\i}az - Sol\'orzano.}
\email{srafael@ipc.upel.edu.ve}
\affiliation{Departamento de F\'{\i}sica, Universidad Sim\'on Bol\'{\i}var, Sartenejas, Edo. Miranda 89000,
Venezuela.}
\affiliation{Centro de Investigaciones de Matem\'atica y F\'{\i}sica, Departamento de Matem\'atica y F\'{\i}sica, Instituto Pedag\'ogico de Caracas, UPEL, Av. P\'aez, Caracas 1021, Venezuela.}

\begin{abstract}
Physical self - adjoint extensions and their spectra of the simplest one - dimensional Hamiltonian operator in which the mass is constant except for a finite jump at one point of the real axis are correctly found. Some self - adjoint extensions are used to model different kinds of semiconductor heterojunctions within the effective - mass approximation. Their properties and relation to different boundary conditions on envelope wave functions are studied. The limiting case of equal masses (with no mass jump) are reviewed.
\end{abstract}

\maketitle

\section{Introduction}\label{intro}
A contemporary issue in the physics literature concerns finding the appropriate Hamiltonian operator (functional form and
domain) that arises from the application of the effective - mass
approximation to semiconductor hetero\-structures when the
effective mass is position - dependent and only piecewise continuous
\cite{gadella-matching,gadelladeltawell,gadella-point-mass-jump}.
In the effective - mass model the momentum operator no longer
conmutes with the mass since the latter depends on position,
therefore the generalization of the standard Hamiltonian operator
is not trivial. The basic problems are the following: a) the
choice of a correct ordering of the mass and momentum operators
together with the boundary conditions on the wave functions across
an abrupt heterojunction to build the Hamiltonian operator so as
to be self-adjoint, and b) the Galilean invariance related to the
Bargmann's theorem \cite{hagen}. According to
\cite{levy-galilean}, the Hamiltonian operator arising from the
application of the effective - mass model that fulfills Galilean
invariance is
\begin{equation}\label{galilean}
H=\frac{1}{2}P\frac{1}{m}P.
\end{equation}
In particular, if $m$ is constant the expression $H=\frac{P^{2}}{2m}$ is recovered.\\
\par\indent
Physical systems with an abrupt discontinuity of the mass at one point model the behavior of a quantum
particle, i.e. an electron moving in a medium formed by two different materials. On each material
the particle behaves as if it had a different mass. The discontinuity point represents the junction between
these two materials.\\
\par\indent
The simplest model, which has been studied by many authors, in an unclear and incomplete form, is given by a one - dimensional system in which the mass is constant except for a finite jump at one point of the real axis, which is chosen to be the origin for simplicity,
\begin{equation}\label{piecemass}
m(x)=\begin{cases}
m_{l} &\textrm{if}\hspace{.2cm} x<0,\\
m_{r} &\textrm{if}\hspace{.2cm} x>0,
\end{cases}
\end{equation}
where $m_{l}\neq m_{r}$ are positive constants. In this case, the Hamiltonian operator has the functional form
\begin{equation}\label{piecehamiltonian}
H=\begin{cases}
-\frac{\hslash^2}{2m_{l}}\frac{d^{2}}{dx^{2}} &\textrm{if}\hspace{.2cm} x<0,\\*[.2cm]
-\frac{\hslash^2}{2m_{r}}\frac{d^{2}}{dx^{2}} &\textrm{if}\hspace{.2cm} x>0
\end{cases}
\end{equation}
In a suitable domain, \eqref{piecehamiltonian} has infinite self-adjoint extensions \cite{gadella-matching}.
\par\indent
In this paper, we show that all the self-adjoint extensions have discrete spectrum. We examine which extensions could play an interesting role according to physical arguments.\\
\par\indent
The paper is organised as follows: in section \ref{extensions}, we find the set of all possible self-adjoint
extensions of $H$, and we use physics constraints to reduce the set. In section \ref{spectra}, we
calculate the reflection and transmission coefficients for all self - adjoint extensions that satisfy
physics constraints. From the equation of the poles of the scattering coefficients, we obtain the
equation that characterizes the spectrum of each physical self-adjoint extension.

\section{Self - adjoint extensions of H}\label{extensions}
We will follow \cite{simon} and \cite{naimark} to construct the self-adjoint extensions of $H$. To construct
the self - adjoint extensions of the operator $H$ we must begin by defining the smaller domain where it makes
sense operator action. In this section we will assume that the operator $H$ is densely defined, symmetric and
closed and let $H^{\dag}=H$ be its adjoint. The domain of the operator $H$, $\mathcal{D}(H)$, is a subspace of
$L^2(\mathbb{R})$, i.e.,
\begin{equation}\label{domh}
\mathcal{D}(H)=\left\{\phi\in W^2_2(\mathbb{R})\, ,\,\phi(0^-)=\phi(0^+)=\phi'(0^-)=\phi'(0^+)=0\right\}
\end{equation}
$W^2_2(\mathbb{R})$ is a Sobolev space.

The domain of $H^{\dag}$ is
\begin{equation}\label{domhdag}
\mathcal{D}(H^\dag)=\left\{\phi\in W^2_2(\mathbb{R}-\{0\})\right\}
\end{equation}
The functions in $W^2_2(\mathbb{R}-\{0\})$ satisfy the same properties of the functions in $W^2_2(\mathbb{R})$ except that they and their derivatives may admit a finite jump at the origin. Note that $\mathcal{D}(H)\subset\mathcal{D}(H^{\dag})$.

The deficiency subspaces of $H$ are given by
\begin{subequations}\label{defecto}
\begin{eqnarray}
\label{defecto1}
\mbox{}\hspace{-.5truecm}\mathcal{N}_{+}&=&
\left\{\psi_+\in\mathcal{D}(H^\dag)\,,\, H^\dag\psi_+=iE_0\psi_+\, ,E_0>0\right\},
\\
\label{defecto2}
\mbox{}\hspace{-.5truecm}\mathcal{N}_{-}&=&
\left\{\psi_{-}\in\mathcal{D}(H^\dag)\,,\, H^\dag\psi_-=-iE_0\psi_-\,,E_0>0\right\},
\end{eqnarray}
\end{subequations}
with the respective dimensions $n_{+}$, $n_{-}$. These are called the deficiency indices of the operator $H$
and will be denoted by the ordered pair $\left(n_{+},n_{-}\right)$. The normalized solutions of
$H^\dag\psi_\pm=\pm iE_0\psi_\pm$ are
\begin{subequations}\label{solsk}
\begin{eqnarray}
\label{solskmasmas}
\psi_+^{(+)}(x)&=&\left[\tfrac{4m_rE_0}{\hbar^2}\right]^{1/4}\theta(x)\,e^{-\frac{\sqrt{m_rE_0}}{\hbar}(1-i)x},\\
\label{solskmasmen}
\psi_+^{(-)}(x)&=&\left[\tfrac{4m_lE_0}{\hbar^2}\right]^{1/4}\theta(-x)\,e^{\frac{\sqrt{m_lE_0}}{\hbar}(1-i)x},\\
\label{solskmenmas}
\psi_-^{(+)}(x)&=&\left[\tfrac{4m_rE_0}{\hbar^2}\right]^{1/4}\theta(x)\,e^{-\frac{\sqrt{m_rE_0}}{\hbar}(1+i)x},\\
\label{solskmenmen}
\psi_-^{(-)}(x)&=&\left[\tfrac{4m_lE_0}{\hbar^2}\right]^{1/4}\theta(-x)\,e^{\frac{\sqrt{m_lE_0}}{\hbar}(1+i)x},
\end{eqnarray}
\end{subequations}
where $\theta(x)$ represents the Heaviside step function. Since all the solutions of equations
$H^\dag\psi_\pm=\pm iE_0\psi_\pm$ belong to $L^2(\mathbb{R})$, the deficiency indices are $(2,2)$ and, the
according to Naimark \cite{naimark}, every self - adjoint extensions are parametrized by a $U(2)$ matrix.
This matrix defines a unique self - adjoint extension, $H_{U}$, of $H$ with domain characterized by means
of the set of all functions $\phi\in\mathcal{D}(H^\dag)$ which satisfy the conditions
\begin{equation}\label{conditions}
\frac{1}{m_l}W\left[\bar{\psi}_{\mu},\phi;0^{-}\right]=\frac{1}{m_r}W\left[\bar{\psi}_{\mu},\phi;0^{+}\right],\;\mu=1,2,
\end{equation}
where $W[\bar{\psi},\phi;x]$ is the Wronskian of the functions $\bar{\psi}(x)$ (the bar represents complex conjugate) and $\phi(x)$ at the point $x$, and
\begin{equation}\label{basemasmen}
\begin{bmatrix}\psi_1(x) \\*[.2cm] \psi_2(x)\end{bmatrix}=
\begin{bmatrix}\psi_+^{(+)}(x) \\ \psi_+^{(-)}(x)\end{bmatrix}+
\begin{bmatrix}U_{11} & U_{12} \\*[.2cm] U_{21} & U_{22}\end{bmatrix}
\begin{bmatrix}\psi_-^{(+)}(x) \\ \psi_-^{(-)}(x)\end{bmatrix}
\end{equation}
We will denote the matrix $\left[\begin{smallmatrix}
U_{11} & U_{12}\\U_{21} & U_{22}\end{smallmatrix}\right]$ by $\mathbb{U}$. The expressions
\eqref{conditions} can be written in the form
\begin{equation}\label{boundc1}
\begin{bmatrix}\phi(0^+) \\ \phi'(0^+)\end{bmatrix}
=\mathbb{T}\begin{bmatrix}\phi(0^-) \\ \phi'(0^-)\end{bmatrix},
\end{equation}
where $\phi(0^\pm)\equiv\lim\limits_{x\rightarrow 0^{\pm}}\phi(x)$,
$\phi'(0^\pm)\equiv\lim\limits_{x\rightarrow 0^{\pm}}\phi'(x)$, and the $n_{+}\times n_{-}\;$ matrix $\mathbb{T}$ is given by
\begin{equation}\label{matrixT}
\mathbb{T}=\frac{m_r}{m_l}
\begin{bmatrix}
-\bar{\psi}'_1(0^+) & \bar{\psi}_1(0^+)\\
-\bar{\psi}'_2(0^+) & \bar{\psi}_2(0^+)
\end{bmatrix}^{-1}
\begin{bmatrix}
\bar{\psi}'_1(0^-) & -\bar{\psi}_1(0^-)\\
\bar{\psi}'_2(0^-) & -\bar{\psi}_2(0^-)
\end{bmatrix}
\end{equation}
The matrix $\mathbb{T}$ gives the matching conditions at the origin. From \eqref{basemasmen}, we can rewrite
the matrix $\mathbb{T}$ in the form
\begin{equation}\label{matrixTu}
\mathbb{T}=
\begin{bmatrix}
-\frac{(1+i)[\det(\bar{U})+\bar{U}_{22}+i(\bar{U}_{11}+1)]m_r^{1/4}}{2\bar{U}_{12}m_l^{1/4}} & \frac{i[\det(\bar{U})+1+\bar{U}_{11}+\bar{U}_{22}]m_r^{1/4}}{\sqrt{2}\bar{U}_{12}m_l^{3/4}\sqrt{\frac{E_0}{\hbar^2}}}\\
\frac{[\det(\bar{U})-1+i(\bar{U}_{11}+\bar{U}_{22})]m_r^{3/4}\sqrt{\frac{E_0}{\hbar^2}}}{\sqrt{2}\bar{U}_{12}m_l^{1/4}} & -\frac{(1+i)[\det(\bar{U})+\bar{U}_{11}+i(\bar{U}_{22}+1)]m_r^{3/4}}{2\bar{U}_{12}m_l^{3/4}}
\end{bmatrix}
\end{equation}
The determinant of \eqref{matrixTu} is given by
\begin{equation}\label{detmatrixTu}
\det(\mathbb{T})=\frac{m_r\bar{U}_{21}}{m_l\bar{U}_{12}}.
\end{equation}
From \eqref{conditions}, we obtain that the matrix $\mathbb{T}$ satisfies the expression
\begin{equation}\label{condSobreT}
\mathbb{T}^\dag(i\sigma_y)\mathbb{T}=\frac{m_r}{m_l}(i\sigma_y),
\end{equation}
where $\sigma_y$ is a Pauli's matrix. By comparing \eqref{detmatrixTu} with \eqref{condSobreT}, we have that $|U_{12}|=|U_{21}|$. However, the time reversal invariance ensures that $U_{12}=U_{21}$, as will be shown below.

The time reversal invariance of the Schr\"{o}dinger equation
$$i\hslash\frac{\partial\Psi}{\partial t}(x,t)=H\Psi(x,t)$$
means that if $\Psi(x,t)$ is a solution of the equation, then $\bar{\Psi}(x,t)$, with $t$ replaced by $-t$, is also a solution. If $$\Psi(x,t)=\phi_{E}(x)e^{-i\frac{E}{\hslash}t},$$
the previous statement implies that $\phi_{E}(x)$ and $\bar{\phi}_{E}(x)$ are two eigenfunctions of the Hamiltonian $H$ with the same eigenvalue $E$. The shortcoming for argument is that the boundary conditions \eqref{boundc1}, with $\mathbb{T}$ given by \eqref{matrixTu}, do not lead necessarily to real eigenfunctions $\phi_{E}(x)$. Among all of the self-adjoint extensions of the Hamiltonian only one subclass will have real eigenfunctions. These extensions will be said to be time reversal invariant \cite{sakurai}. The reality of $\phi_{E}(x)$ implies that $\bar{\mathbb{T}}=\mathbb{T}$, i.e., $\mathbb{T}$ is a real matrix. Thus, from \eqref{condSobreT}, we obtain
\begin{equation}\label{detmatrixTreal}
\det(\mathbb{T})=\frac{m_r}{m_l},
\end{equation}
which is mentioned in \cite{ando}. By comparing \eqref{detmatrixTu} with \eqref{detmatrixTreal}, we have
that $U_{12}=U_{21}\neq0$. In conclusion, the reality of the matrix $\mathbb{T}$ makes the Hamiltonian
\eqref{piecehamiltonian} with domain \eqref{boundc1} invariant under time reversal.

\section{Scattering coefficients and the spectra of H}\label{spectra}
In this section we will derive the spectra for the self - adjoint extensions $H_{U}$ from the poles of scattering
amplitudes. For this, let us parametrize the unitary matrix $\mathbb{U}$ as
\begin{equation}\label{Uparametrizada}
\mathbb{U}=e^{i\psi}\mathbb{A},\;
\det(\mathbb{A})=1,
\end{equation}
where
\begin{equation}\label{matrixA}
\mathbb{A}=\begin{bmatrix}
  a_0-ia_3 & -a_2-ia_1 \\
  a_2-ia_1 & a_0+ia_3 \\
\end{bmatrix},
\end{equation}
with $a_0,a_1,a_2,a_3\in\mathbb{R}$, and $\psi\in\left[0,\pi\right]$. Notice that the points $\psi=0$ and
$\psi=\pi$ have to be identified. The condition $U_{12}=U_{21}\neq0$ implies $a_2=0$ and $a_1\neq0$. Thus,
by substituting \eqref{Uparametrizada} and \eqref{matrixA} in \eqref{matrixT}, we obtain the real matrix
\begin{equation}\label{matrixTa}
\mathbb{T}=\begin{bmatrix}
-\frac{\left[a_0-a_3+\cos\psi-\sin\psi\right]m_r^{1/4}}{a_1m_l^{1/4}} & \frac{\hslash[a_0+\cos\psi]m_r^{1/4}}{a_1m_l^{3/4}\sqrt{E_0}}\\
 \frac{2\sqrt{E_0}[a_0-\sin\psi]m_r^{3/4}}{a_1m_l^{1/4}\hslash} & -\frac{\left[a_0+a_3+\cos\psi-\sin\psi\right]m_r^{3/4}}{a_1m_l^{3/4}}
\end{bmatrix}
\end{equation}
In terms of \eqref{matrixTa}, the matching conditions \eqref{boundc1} are
\begin{equation}\label{bounconda}
\begin{bmatrix}\phi(0^+) \\*[.4cm] \phi'(0^+)\end{bmatrix}=
\begin{bmatrix}
-\frac{\left[a_0-a_3+\cos\psi-\sin\psi\right]m_r^{1/4}}{a_1m_l^{1/4}} & \frac{\hslash\left[a_0+\cos\psi\right]m_r^{1/4}}{a_1m_l^{3/4}\sqrt{E_0}}\\
 \frac{2\sqrt{E_0}\left[a_0-\sin\psi\right]m_r^{3/4}}{a_1m_l^{1/4}\hslash} & -\frac{\left[a_0+a_3+\cos\psi-\sin\psi\right]m_r^{3/4}}{a_1m_l^{3/4}}
\end{bmatrix}\,
\begin{bmatrix}\phi(0^-) \\*[.4cm] \phi'(0^-)\end{bmatrix}
\end{equation}
\indent
Let us assume that an incoming monochromatic wave $e^{ik_{l}}$, $k_{l}=\sqrt{\frac{2m_{l}E}{\hslash^2}}$, $E>0$,
comes from the left, so that the wave function for $x<0$ is $e^{ik_{l}}+r_l e^{-ik_{l}}$, and the wave function
for $x>0$ is $t_le^{ik_{r}}$, $k_{r}=\sqrt{\frac{2m_{r}E}{\hslash^2}}$, $E>0$, where $r_l$ and $t_l$ are the
reflection and transmission amplitudes, respectively, for incoming wave from the left. Then, the matching
conditions \eqref{bounconda} at the origin give
\begin{equation}\label{condRT}
\begin{bmatrix}1+r_l \\*[.4cm] ik_{l}(1-r_l)\end{bmatrix}=
\begin{bmatrix}
-\frac{\left[a_0-a_3+\cos\psi-\sin\psi\right]m_r^{1/4}}{a_1m_l^{1/4}} & \frac{\hslash\left[a_0+\cos\psi\right](2m_r)^{1/4}}{a_1m_l^{3/4}\sqrt{E_0}}\\
 \frac{2\sqrt{E_0}\left[a_0-\sin\psi\right]m_r^{3/4}}{a_1m_l^{1/4}\hslash} & -\frac{\left[a_0+a_3+\cos\psi-\sin\psi\right]m_r^{3/4}}{a_1m_l^{3/4}}
\end{bmatrix}\,
\begin{bmatrix}t_l \\*[.5cm] ik_{r}t_l\end{bmatrix}
\end{equation}
and then one finally obtains the expressions of $r_l$ and $t_l$ as
\begin{subequations}
\begin{eqnarray}
\label{r}
r_l&=&\tfrac{E_0(a_0-\sin\psi)+E(a_0+\cos\psi)-i \sqrt{2} a_3 \sqrt{E} \sqrt{E_0}}{E(a_0+\cos\psi)-E_0 (a_0-\sin\psi) -i \sqrt{2}\sqrt{E}\sqrt{E_0} (a_0-\sin\psi+\cos\psi)}\\
\label{t}
t_l&=&\tfrac{i\sqrt{2}a_1 \sqrt{E} \sqrt{E_0} \sqrt[4]{m_r}}{\sqrt[4]{m_l} \left(E(a_0+\cos\psi)-E_0 (a_0-\sin\psi) -i \sqrt{2}\sqrt{E}\sqrt{E_0} (a_0-\sin\psi+\cos\psi)\right)}
\end{eqnarray}
\end{subequations}
Making use of $a_0^2+a_1^2+a_3^2=1$, we have $\left|r_l\right|^2+\left|t_l\right|^2\sqrt{\frac{m_l}{m_r}}=1$.
The poles of $r_l$ and $t_l$ satisfy the following equation
\begin{equation}\label{poles}
E(a_0+\cos\psi)-E_0 (a_0-\sin\psi) -i \sqrt{2}\sqrt{E}\sqrt{E_0} (a_0-\sin\psi+\cos\psi)=0
\end{equation}
The poles of $r_r$ and $t_r$ ($r_r$ and $t_r$ are the reflection and transmission amplitudes, respectively,
for incoming wave come from the right) also satisfy \eqref{poles}.\\
\par\indent
The zero values of \eqref{r} correspond to transmission resonances \cite{newton}.
The roots of \eqref{poles} can be written as
\begin{alignat}{2}
\label{energybound1}
\sqrt{E_{\pm}}&=i\left(1\pm\frac{\sqrt{1-a_0^2}\mp\sin\psi}{a_0+\cos\psi}\right)\sqrt{\frac{E_0}{2}},
&\hspace{.5cm}(\textrm{if}\;a_0\neq-\cos\psi)\\
\label{energybound2}
\sqrt{E}&=i\sqrt{\frac{E_0}{2}}\left(1+\cot\psi\right),
&(\textrm{if}\;a_0=-\cos\psi)
\end{alignat}
Note that $E_{\pm}$ and $E$ are real for each self - adjoint extension $H_U$. If the expressions within the parentheses in \eqref{energybound1} and \eqref{energybound2} are positive (negative), then the square root of energy is in the physical Riemann sheet (unphysical Riemann sheet) \cite{newton,taylor}. If the square root of energy is positive, then we have a bound state, otherwise, we have an antibound (virtual) state. Some self - adjoint extensions have two different values of energy.\\
\par\indent
In the next subsections, we discuss the spectrum of some self-adjoint extension of \eqref{piecehamiltonian}
corresponding to one - dimensional Hamiltonian: (1) with a delta potential at the origin plus mass jump at the
same point, (2) with a delta plus delta prime potential at the origin plus mass jump at the same point and, (3)
with a delta prime potential at the origin plus mass jump at the same point.

\subsection{Hamiltonian with a delta potential at the origin plus a mass jump at the same point}\label{Hconpoint}
The boundary conditions corresponding to one - dimensional Hamiltonian with a delta potential at the origin
plus a mass jump at the same point \cite{alavrezresonanceswithdelta} are
\begin{equation}\label{deltawithjump}
\begin{bmatrix} \phi(0^+)\\\phi'(0^+)\end{bmatrix}=
\begin{bmatrix}
1 & 0\\
\frac{2m_r\gamma}{\hbar^2} & \frac{m_r}{m_l}
\end{bmatrix}
\begin{bmatrix} \phi(0^-)\\\phi'(0^-)\end{bmatrix},
\end{equation}
where $\gamma$ is the strength of delta potential. For the square root of energy being in the physical Riemann
sheet, the parameter $\gamma$ must be negative. By comparing \eqref{deltawithjump} with \eqref{matrixTa}, we
obtain the following values of the parameters $a_0$, $a_1$ and $a_3$
\begin{subequations}\label{aes-delta}
\begin{align}
a_0&=-\cos\psi,\\
a_1&=-\frac{\sqrt{E_0}\hslash(\cos\psi+\sin\psi)}{(m_lm_r)^{1/4}\gamma},\\
a_3&=-\frac{\sqrt{E_0}\hslash(\cos\psi+\sin\psi)+\sqrt{m_r}\gamma\sin\psi}{\sqrt{m_r}\gamma}
\end{align}
\end{subequations}
Using the constraint $a_0^2+a_1^2+a_3^2=1$, we obtain
\begin{equation}\label{gamma}
1+\cot\psi=\frac{2\sqrt{m_rm_l}(-\gamma)}{\hslash\sqrt{E_0}(\sqrt{m_l}+\sqrt{m_r})}
\end{equation}
Then, by inserting \eqref{gamma} in \eqref{energybound2}, we have
\begin{equation}\label{deltawithjumpenergy}
E=-2\frac{m_l m_r\gamma^2}{\hslash^2\left(\sqrt{m_l}+\sqrt{m_r}\right)^2}
\end{equation}
The energy of the unique bound state is always negative (except if $\gamma=0$, in which no bound state is
present). This eigenvalue is unique. Thus, the one - dimensional Hamiltonian with a delta potential at the origin
plus a mass jump at the same point has a bound state, unlike it was stated in \cite{gadelladeltawell}, that the
mass jump cannot exist with the term $-\gamma\delta(x)$ only, unless the term $\lambda\delta'(x)$ is present.

\subsection{Hamiltonian with a delta plus a delta prime potential at the origin plus a mass jump at the same point}\label{Hconpoint-prime}
The matching conditions for this self -adjoint extension are
\begin{equation}\label{bouncond-delta-deltaprime}
\begin{bmatrix}\phi(0^+) \\*[.5cm] \phi'(0^+)\end{bmatrix}=
\begin{bmatrix}
\frac{m_r}{m_l}\frac{\hslash^2 (m_l+m_r)+2\lambda m_l m_r}{\hslash^2 (m_l+m_r)-2\lambda m_l m_r}& 0\\
 \frac{2 \gamma m_r \left(\hslash^2 (m_r+m_l)^2+2 \lambda m_l
  m_r (m_r-m_l)\right)}{\hslash^4 (m_l+m_r)^2-4 \lambda ^2
   m_l^2 m_r^2}& \frac{\hslash^2 (m_l+m_r)-2\lambda m_l m_r}{\hslash^2 (m_l+m_r)+2\lambda m_l m_r}
\end{bmatrix}\,
\begin{bmatrix}\phi(0^-) \\*[.5cm] \phi'(0^-)\end{bmatrix},
\end{equation}
where $\lambda$ is the strength of delta prime potential. The matching conditions are obtained in
\cite{gadelladeltawell} for $\alpha=-\beta=-1$. The determinant of the matrix $\mathbb{T}$ in
\eqref{bouncond-delta-deltaprime} satisfies identically \eqref{detmatrixTreal} without any condition on
$\lambda$, unlike the statement in \cite{gadelladeltawell}.\\
\par\indent
If $m_l=m_r\equiv m$ in \eqref{bouncond-delta-deltaprime}, we obtain the matching conditions at the origin
for the Hamiltonian with no mass jump with a delta plus delta prime potential \cite{kurasov}. Note that
\eqref{bouncond-delta-deltaprime} is reduced to the boundary conditions \eqref{deltawithjump} if
$\lambda=\frac{\hslash^2}{2m_lm_r}(m_l-m_r)$. In this case, the value $\lambda=0$ leads to Hamiltonian with no
mass jump ($m_l=m_r\equiv m$) with a delta potential \cite{kurasov}.\\
\par\indent
Again, the energy of the bound state is always negative (except if $\gamma=0$ or $\lambda=\hslash^2\frac{ \left(m_l+m_r\right)^2}{2m_lm_r(m_l-m_r)}$, in which no bound state is present)
\begin{equation}\label{delta-plus-prime-energy}
E=-\frac{2\gamma^2m_l^2m_r^2}{\hbar^2}
\left[
\frac{1-\frac{2\mu\lambda}{\hbar^2}\left(\frac{m_l-m_r}{m_l+m_r}\right)}{(1-\frac{2\mu\lambda}{\hbar^2})^2m_l^{3/2}+(1+\frac{2\mu\lambda}{\hbar^2})^2m_r^{3/2}}
\right]^2,
\end{equation}
where $\mu=m_lm_r/(m_l+m_r)$. For the square root of energy being in the physical Riemann sheet, we need that
$\gamma<0$ and $1-\frac{2\mu\lambda}{\hbar^2}\left(\frac{m_l-m_r}{m_l+m_r}\right)>0$ or that $\gamma>0$ and
$1-\frac{2\mu\lambda}{\hbar^2}\left(\frac{m_l-m_r}{m_l+m_r}\right)<0$. Using $\lambda=0$ in
\eqref{delta-plus-prime-energy}, we obtain
\begin{equation}\label{delta-plus-prime-energy-lambda-cero}
E_{\lambda=0}=-2\frac{m_l^2 m_r^2\gamma ^2}{\hslash^2 \left(m_l^{3/2}+m_r^{3/2}\right)^2}
\end{equation}
This energy corresponds to the self - adjoint extension whose matching conditions at the origin are
\begin{align}\label{boundc-delta-plus-prime-lambda-cero}
\begin{bmatrix}\phi(0^+) \\ \phi'(0^+)\end{bmatrix}=
\begin{bmatrix}
\frac{m_r}{m_l}& 0\\
 \frac{2 \gamma m_r}{\hslash^2}&1
\end{bmatrix}\,
\begin{bmatrix}\phi(0^-) \\ \phi'(0^-)\end{bmatrix}
\end{align}
This self - adjoint extension corresponds to a Hamiltonian with mass jump plus a delta potential at the origin.
Mass jump form is different to that shown in \eqref{deltawithjump}. Unlike it was stated in
\cite{gadelladeltawell}, the presence of a delta prime potential is not required in the Hamiltonian for the
jump mass can exist.\\
\par\indent
If $m_l=m_r\equiv m$, both \eqref{deltawithjumpenergy} and \eqref{delta-plus-prime-energy-lambda-cero} reduce
to
\begin{equation*}
E=-\frac{m\gamma^2}{2\hslash^2}
\end{equation*}
This energy corresponds to the bound state energy of the Hamiltonian no mass jump with a delta potential
\cite{albeverio}.

\subsection{Hamiltonian with a delta prime potential at the origin plus a mass jump at the same point}\label{Hconpointprime-only}
The boundary conditions corresponding to one - dimensional Hamiltonian with a delta prime potential at the
origin plus a mass jump at the same point are
\begin{align}\label{boundc-delta-prime}
\begin{bmatrix}\phi(0^+) \\ \phi'(0^+)\end{bmatrix}=
\begin{bmatrix}
\frac{m_r}{m_l}&\frac{\sqrt{2}m_r\lambda}{\hslash\sqrt{m_l}\sqrt{E_0}}\\
 0&1
\end{bmatrix}\,
\begin{bmatrix}\phi(0^-) \\ \phi'(0^-)\end{bmatrix},\,\lambda<0
\end{align}
These boundary conditions are similar to those in the case of equal masses \cite{kurasov,albeverio}, including
$\lambda<0$. The energy of the unique bound state is always negative (except if $\lambda=0$, in which no bound
state is present). This eigenvalue is unique
\begin{equation}\label{delta-prime-energy}
E=-\frac{E_0\hslash^4}{4}\frac{\left(m_l^{3/2}+m_r^{3/2}\right)^2}{m_l^2m_r^3\lambda^2}
\end{equation}

If $m_l=m_r\equiv m$, \eqref{delta-prime-energy} reduces to the bound state energy of the Hamiltonian with no mass
jump with a delta prime potential \cite{albeverio}, i.e.,
\begin{equation*}
E=-\frac{E_0\hslash^4}{m^2\lambda^2}
\end{equation*}

It is interesting to remark that if $\gamma=0$ in \eqref{bouncond-delta-deltaprime}, we obtain a self - adjoint extension corresponding to the one-dimensional Hamiltonian with a mass jump and dipole point interaction (first derivative of the delta potential), which has no bound states. For $\lambda=\pm\hslash^2\frac{m_l+m_r}{2m_l m_r}$, we have that the transmission coefficient is zero, whereas for the lambda values
$\frac{(m_l+m_r)\left(m_l^{\frac{3}{4}}\pm m_r^{\frac{3}{4}}\right)^2}{2m_l m_r \left(m_l^{\frac{3}{2}}-m_r^{\frac{3}{2}}\right)}$, the transmission coefficient is one (we have a transmission resonance). Thus, we have two different physical self - adjoint extension despite they have the same functional form. It should be noted that in the case of equal masses the derivative of the delta potential has not bound states \cite{kurasov}.

\section{Concluding remarks.}\label{remark}
Using the Von Neumann's theory of self - adjoint extensions and physical arguments, we found the general
matching conditions \eqref{bounconda} that describe each one of the different domains of the various
self - adjoint extensions of \eqref{piecehamiltonian}. Some boundary conditions have been reported
\cite{gadelladeltawell,alavrezresonanceswithdelta,levy-elementary}. Using scattering theory, we obtained
the spectrum of each one of the extensions. Each physical self - adjoint extensions correspond to one different Hamiltonian operator. Given that the operator \eqref{piecehamiltonian} has self - adjoint extensions with bound states only, it is not appropriate to call the operator \eqref{piecehamiltonian} a kinetic Hamiltonian.
\par\indent
Finally, we analyze three different self - adjoint extensions of \eqref{piecehamiltonian}. We discuss the
limiting case $m_l=m_r\equiv m$ for each one of the three self - adjoint extensions. In this limit, the
self - adjoint extensions above mentioned correspond to respective point interaction extension
\cite{kurasov,albeverio}.

\section{Acknowledgments}

This work was supported by IVIC under project 1089. The authors would like to thank the referee for the positive
and valuable suggestions, and the Drs. Jhoan Toro and Ernesto Medina for reading and improving the manuscript.

\end{document}